\documentclass[%
aps,pra,%
amsmath,amssymb,
reprint,%
floatfix
]{revtex4-1}
\usepackage[utf8]{inputenc}
\usepackage[english]{babel}
\usepackage[colorlinks=true]{hyperref}
\usepackage{siunitx}
\usepackage{amsmath,amssymb}
\usepackage{graphicx}
\usepackage{xcolor}

\usepackage[english,plain]{fancyref}

\frefformat{plain}{\fancyreffiglabelprefix}{Fig.~#1}
\frefformat{plain}{\fancyrefeqlabelprefix}{Eq.~#1}
\frefformat{plain}{\fancyreftablabelprefix}{TABLE~#1}

\usepackage{xfrac}

\begin{document}
    \makeatletter
    \title{Optical amplification of spin noise spectroscopy via homodyne detection}
    \let\theTitle\@title
    \makeatother

    \author{Pavel Sterin}
    \affiliation{Institut f\"{u}r Festk\"{o}rperphysik, Leibniz Universit\"{a}t Hannover, Appelstr.~2, D-30167 Hannover, Germany}

    \author{Julia Wiegand}
    \affiliation{Institut f\"{u}r Festk\"{o}rperphysik, Leibniz Universit\"{a}t Hannover, Appelstr.~2, D-30167 Hannover, Germany}

    \author{Jens H\"{u}bner}
    \affiliation{Institut f\"{u}r Festk\"{o}rperphysik, Leibniz Universit\"{a}t Hannover, Appelstr.~2, D-30167 Hannover, Germany}

    \author{Michael Oestreich}
    \email{oest@nano.uni-hannover.de}
    \affiliation{Institut f\"{u}r Festk\"{o}rperphysik, Leibniz Universit\"{a}t Hannover, Appelstr.~2, D-30167 Hannover, Germany}

\begin{abstract}
Spin noise (SN) spectroscopy measurements on delicate semiconductor spin systems, like single InGaAs quantum dots, are currently not limited by optical shot noise but rather by the electronic noise of the detection system.
Here, we report a realization of homodyne SN spectroscopy enabling shot noise limited SN measurements.
The proof-of-principle measurements on impurities in an isotopically enriched rubidium atom vapor show that homodyne SN spectroscopy can be utilized even in the low frequency spectrum which facilitates advanced semiconductor spin research like higher order SN measurements on spin qubits.
\end{abstract}

\maketitle

\section{Introduction}
Traditional optical experiments study the spin dynamics in semiconductors by optical excitation of spin polarized carriers like in polarization resolved photoluminescence or pump probe Faraday rotation experiments~\cite{meier_optical_1984, kikkawa_resonant_1998}.
However, these techniques are problematic in spin systems where the perturbation by optical excitation changes the intrinsic carrier spin dynamics. This applies to weakly interacting and few particle spins~\cite{Atature.NatPhys.2007, Dahbashi.APL.2012, Dahbashi.PRL.2014} as well as spin systems where additional interaction is induced by optically injected carriers~\cite{Romer.RSI.2007, henn_hot_2013}. For such fragile semiconductor spin systems, spin noise (SN) spectroscopy has been established as the method of choice in order to avoid unnecessary optical excitation~\cite{hubner_rise_2014}.
The experimental technique has been transferred from quantum optics to semiconductor physics in 2005~\cite{oestreich_spin_2005} and is nowadays used to study, e.g., the spin dynamics of single carriers and trions in single InGaAs quantum dots~\cite{dahbashi_optical_2014,wiegand_reoccupation_2017} or the complex interaction of localized electrons with nuclear spins~\cite{berski_interplay_2015}.
In this context SN spectroscopy is either used as weakly disturbing measurement method at thermal equilibrium~\cite{muller_semiconductor_2010} or to study quantum spin systems which are strongly driven by a resonant light field~\cite{horn_spin-noise_2011, wiegand_reoccupation_2017}.
In the first case, very low light intensities are required~\cite{Dahbashi.APL.2012, Dahbashi.PRL.2014} and in the second case, SN at high frequencies~\cite{Muller.PRB.2010, Berski.PRL.2013} and high sensitivity~\cite{wiegand_reoccupation_2017} is of special interest.
In both scenarios, the dominant experimental noise source is not SN or optical shot noise but extrinsic electronic noise from the electro-optical detectors entailing long measuring times. This constraint \emph{inter alia} impedes fundamental measurements like high-frequency single spin dynamics or the efficient measurement of higher correlation SN.

\begin{flushright}
\begin{figure}[t]
 \centering
 \includegraphics[]{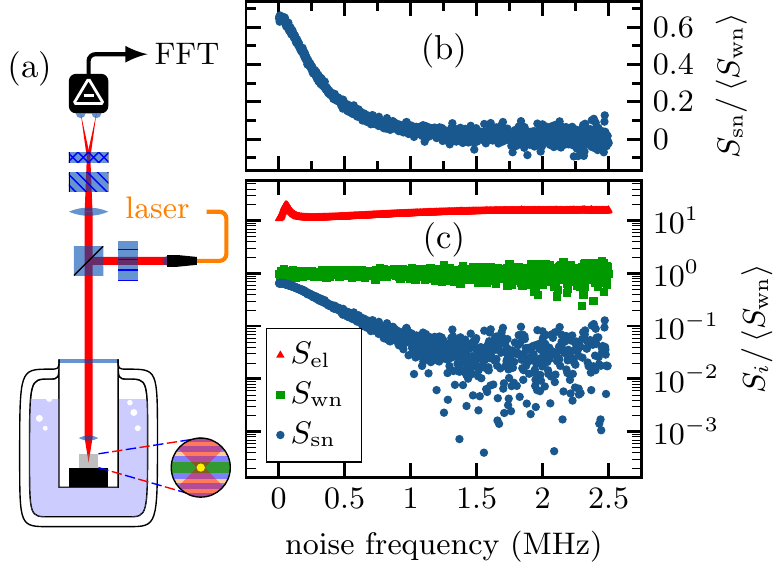}
 \caption{(a) Schematic of a typical SN spectroscopy setup for single quantum dot SN measurements in reflection (see Ref.~\onlinecite{wiegand_reoccupation_2017} for details).
 (b) Typical SN power spectrum of a single InGaAs quantum dot charged by a hole in units of the probe laser shot noise power.
 (c) Comparison on a logarithmic scale of the broadband electronic noise of the balanced receiver $S_{\rm en}$, the white photon shot noise $S_{\textrm{wn}}$, and the SN $S_{\textrm{sn}}$ for a typical quantum dot SN measurement.%
}
 \label{fig:spin_noise}
\end{figure}
\end{flushright}

The impact of electronic noise can be diminished by optical amplification.
The technique is known in quantum optics as homodyne detection and we will show in this publication that homodyne detection~\cite{laforge_noninvasive_2007, laforge_machzehnder_2008,cronenberger_quantum_2016} can be transferred smoothly to a standard SN spectroscopy setup even in the regime of \emph{low} noise frequencies.
For high frequencies,  Cronenberger and Scalbert have recently demonstrated quantum limited heterodyne detection of SN in GaAs~\cite{cronenberger_quantum_2016} in a non-standard SN setup.
Here, we demonstrate the optical amplification of the SN signal on $^{85}$Rb impurities in an isotopically enriched $^{87}$Rb vapor, that is a well characterized reference system~\cite{crooker_spectroscopy_2004, horn_spin-noise_2011, ma_spin_2016}. The same technique will also be applicable to delicate, low frequency semiconductor spin systems like a single, charged InGaAs quantum dot at low temperatures where the electronic noise is a dominating obstacle in fast and precise measurements of the spin dynamics. The setup and data presented in \fref{fig:spin_noise} demonstrate the typical dominance of electronic noise in current state-of-the-art QD SN measurements.
Such low-dimensional semiconductor spin systems with long spin lifetimes are of special interest due to their prospective potential as spin qubits in spin information processing~\cite{loss_quantum_1998, imamoglu_quantum_1999, delteil_realization_2017}.

\section{Methods}
\subsection{Spin Noise Spectroscopy}
Spin noise spectroscopy in semiconductors maps the stochastic spin fluctuations of carriers at thermal equilibrium onto the polarization of laser light that is quasi-resonant to an optical transition~\cite{hubner_rise_2014}.
\Fref{fig:spin_noise}(a) depicts a typical schematic experimental setup for SN measurements on a single semiconductor quantum dot in reflection.
Here, linearly polarized laser light is focused on a single InGaAs quantum dot which is embedded in an asymmetric AlAs/GaAs Bragg resonator and charged by a single hole.
The stochastic spin polarization of the hole yields a Faraday rotation of the linear polarization of the laser light~\cite{hubner_rise_2014} 
which is measured by a polarization bridge and a balanced receiver.
The time-continuous electrical signal is amplified, digitized, and Fourier transformed.
The resulting Fourier power spectra are averaged ($S_\textrm{fg}$) and an averaged background spectrum ($S_\textrm{bg}$) is subtracted.
The background spectrum is measured by applying a transverse magnetic field which shifts the SN to higher frequencies outside of the detection bandwidth  for the measurement depicted in \fref{fig:spin_noise}.
\Fref{fig:spin_noise}(b) shows exemplarily such a single quantum dot SN spectrum for a laser intensity of \SI[per-mode=symbol]{1.1}{\micro\watt\per\micro\meter\squared} while \fref{fig:spin_noise}(c) compares the corresponding noise contributions of the electronic noise ($S_\textrm{en}$, red dots), the white optical shot noise ($S_\textrm{wn}$, green dots), and the weak SN ($S_\textrm{sn}$, blue dots).
The dominant electronic noise results mainly from the commercial low noise balanced photo-receiver 
which is in this typical SN QD experiment about one order of magnitude larger than the optical shot noise, i.e., the electronic noise increases the necessary measurement time by two orders of magnitude.
Intensity dependent measurements (not shown) reveal that nevertheless the depicted SN spectrum is strongly influenced by the light field despite the rather low probe laser intensity~\cite{dahbashi_optical_2014}.
In fact, orders of magnitude lower laser intensities are desirable for such SN experiments whereat optical amplification by homodyne detection becomes essential.

\subsection{Homodyne Detection}

In this section we briefly introduce the principle of homodyne amplification in the context of standard SN spectroscopy.
Optical homodyne detection utilizes two phase-locked, monochromatic laser beams which can be described as classical electromagnetic plane waves.
For SN spectroscopy only the electric fields and the polarization states of the two beams are relevant which both can be represented by two Jones vectors in a linearly polarized basis~\cite{jackson_electrodynamics_2007}.
The vertically polarized signal beam travels through the sample and experiences a small stochastic Faraday rotation by an angle $\theta_\textrm{F}$.
The resulting electric field vector is
\begin{equation}
    \mathbf{E}_\textrm{sig}=
        E_{\textrm{sig},0} R\left(\theta_\textrm{F}\right)\cdot \mathbf{e}_\textrm{V} =
        E_{\textrm{sig},0}\begin{pmatrix}
            \sin\left(\theta_\textrm{F}\right)\\
            \cos\left(\theta_\textrm{F}\right)\\
        \end{pmatrix},
\end{equation}
where $E_{\textrm{sig},0}$ is the complex amplitude, $R\left(\theta_\textrm{F}\right)$ is the regular Euclidean rotation matrix, and $\mathbf{e}_\textrm{V}$ is the polarization vector.
The second beam yields the local oscillator field and acts as unchanged polarization reference
        \begin{equation}
            \mathbf{E}_\textrm{lo}=
                \sqrt{2 I_\textrm{lo}/\epsilon_0 c} \cdot \mathbf{e}_\textrm{V} =
                \sqrt{2 I_\textrm{lo}/\epsilon_0 c} \cdot \begin{pmatrix}
                    0\\
                    1\\
                \end{pmatrix},
        \end{equation}
where $I_\textrm{lo} = p\epsilon_0 c\left|E_{\textrm{sig},0} \right|^2 /2 = p I_\textrm{sig}$ is the intensity of the local oscillator in terms of the intensity of the signal beam and $p$ is the corresponding proportionality factor.
For SN spectroscopy, the interference of both beams $\mathbf{E}_\textrm{if}$ is decomposed into a linearly polarized basis rotated by $\sfrac{\pi}{4}$ (diagonal $\mathbf{e}_\textrm{D}$ and anti-diagonal $\mathbf{e}_\textrm{A}$) giving the components $E_\textrm{if,D}$ and $E_\textrm{if,A}$.
The balanced receiver measures the difference of the intensities of these two components
        \begin{align}
            \Delta I &= {\left|E_\textrm{if,D}\right|^2 - \left|E_\textrm{if,A}\right|^2} \notag \\
            &= 2 I_\textrm{sig} \sin (\theta_\textrm{F}) \left[\cos (\theta_\textrm{F})+\sqrt{p} \cos(\varphi_\textrm{sig,lo} )\right] \notag \\
            &= 2 I_\textrm{sig} \theta_\textrm{F} \left(1 + \sqrt{p} \cos(\varphi_\textrm{sig,lo})\right) + O\left(\theta_\textrm{F}^2\right),
            \label{eq:amp-approx}
        \end{align}
where $\varphi_\textrm{sig,lo}$ is the phase between $\mathbf{E}_\textrm{sig}$ and $\mathbf{E}_\textrm{lo}$ which is typically set to a multiple of $2\pi$ for optimal amplification.
Typical SN measurements involve very small Faraday rotation angles and for this reason the use of small angle approximation is justified. In this case the intensity difference is linear in $\theta_F$ and $(1 + \sqrt{p})$. Finally, the SN signal is proportional to $\Delta I^2$ and consequently scales by the factor $\eta_p = (1+\sqrt{p})^2$. This dependence allows the signal-to-noise ratio of the SN signal to be increased without disturbing the measured system by increasing the local oscillator intensity up to the point where optical shot noise dominates over the electronic noise.
In practice this kind of amplification is limited only by the maximally allowed optical power incident on the photo diodes of the balanced receiver.

\subsection{Noise Power and Signal-to-Noise Ratio}
Spin noise is typically evaluated by the SN power density spectrum $S_\textrm{sn}(f)$ which follows for mono-exponential spin relaxation dynamics a normalized Lorentzian profile of area $1$ multiplied by the SN power amplitude $A$. The Lorentzian is shifted in a transverse magnetic field by the Larmor spin precession frequency $f_\textrm{L}$ and has a width $\gamma_\textrm{s}$ given by the spin dephasing rate:
\begin{equation}
    L_{f_{\textrm{L}},\gamma_\textrm{s}}(f)=\frac{\gamma_\textrm{s}}{\pi \left((f-f_\textrm{L})^2 + \gamma_\textrm{s}^2 \right)}.
\end{equation}
The \textit{measured} SN power, i.e. the integral over $S_\textrm{sn}$, scales with the square of the probe laser power~\cite{hubner_rise_2014} and thus the SN power density scales with the same factor if $\gamma_\textrm{s}$ is constant.
Homodyne detection amplifies  $S_\textrm{sn}$ by approximately $\eta_p$.
Laser photon shot noise yields an additional broadband white noise contribution that scales linearly with the laser power on the photo diodes.
The total laser power at the homodyne operation point of constructive interference ($\varphi_\textrm{sig,lo}=0$) reads
        \begin{equation*}
	            \label{eq:wn_power}
                {P_\textrm{L} = \left(\sqrt{P_\textrm{sig}} + \sqrt{p P_\textrm{sig}}\right)^2 = \eta_p} P_\textrm{sig}\:.
        \end{equation*}
The electronic noise is in general independent of the incident laser power resulting in the total noise power density spectrum
\begin{subequations}
    \label{eq:noise-all}
    \begin{align}
    S(f) = \quad
        &A \;L_{f_{\textrm{L}},\gamma_\textrm{s}} \alpha^2 P_\textrm{sig}^2 \eta_p&
           &\left(\textrm{spin noise}\; S_\textrm{sn}\right)& \label{eq:noise-SN}
           \\
       +\;&W \; \hbar \omega_\textrm{L} \alpha P_\textrm{sig} \eta_p&
           &\left(\textrm{shot noise}\; S_\textrm{wn}\right)& \label{eq:noise-WN}
           \\
       +\;&S_\textrm{en}&
           &\left(\textrm{electronic noise}\right)& \label{eq:noise-EN}
    \end{align}
\end{subequations}
where $W$ is a constant, $\hbar \omega_\textrm{L}$ is the laser photon energy, and $\alpha(f)$ is the (unitless) detector response function.

The signal to noise ratio is given by the spin noise signal $S_\textrm{sn}$ (Eq.~\ref{eq:noise-SN}) divided by the non-signal contributions $S_\textrm{wn}$ (Eq.~\ref{eq:noise-WN}) and $S_\textrm{sn}$ (Eq.~\ref{eq:noise-EN}). Here, however, in order to consistently analyze the performance of homodyne detection we define the signal-to-noise ratio $r_\textrm{SNR}^\star$ as the ratio of the SN amplitude and the spectral mean of the remaining broadband noise contributions:
        \begin{equation}
            \label{eq:snr-star}
            r_\textrm{SNR}^\star =
            \frac{\operatorname{max}(S_\textrm{sn})/P_\textrm{sig}}
            {\langle S_\textrm{en} \rangle + \langle S_\textrm{wn} \rangle}
            = \frac
            {A \; L_{f_\textrm{L},\gamma_\textrm{s}}(f_\textrm{L}) \alpha^2(f_\textrm{L}) \; P_\textrm{L}}
            {\langle S_\textrm{en} \rangle
                + W \; \hbar \omega_\textrm{L} \langle \alpha \rangle \; P_\textrm{L}},
        \end{equation}

where the SN amplitude is divided by the probe laser power $P_\textrm{sig}$ such that $r_\textrm{SNR}^\star$ depends directly on $P_\textrm{L}$ detected on the photo diodes.
This makes $r_\textrm{SNR}^\star$ robust against fluctuations in $P_\textrm{sig}$ and the auxiliary parameters $P_\textrm{sig}$ and $P_\textrm{L}$ can be easily recorded during the measurement. The denominator of $r_\textrm{SNR}^\star$ is the mean of the background spectrum $S_\textrm{bg}$ that is measured by shifting the spin noise outside of the spectral window considered, while the numerator is recovered from the measured difference spectrum ${S_\textrm{sn} \approx S_\textrm{diff} = S_\textrm{fg} - S_\textrm{bg}}$ divided by the measured laser probe power $P_\textrm{sig}$.
This means that $r_\textrm{SNR}^\star$ can be easily estimated in the experiment to

        \begin{equation}
            r_\textrm{SNR}^\star =
                \operatorname{max}\left[\left(S_\textrm{fg} - S_\textrm{bg}\right) / P_\textrm{sig}\right] / \langle S_\textrm{bg} \rangle.
        \end{equation}
\section{Setup}
\subsection{Interferometer}
   \begin{figure}[t]
   \centering
   \includegraphics[]{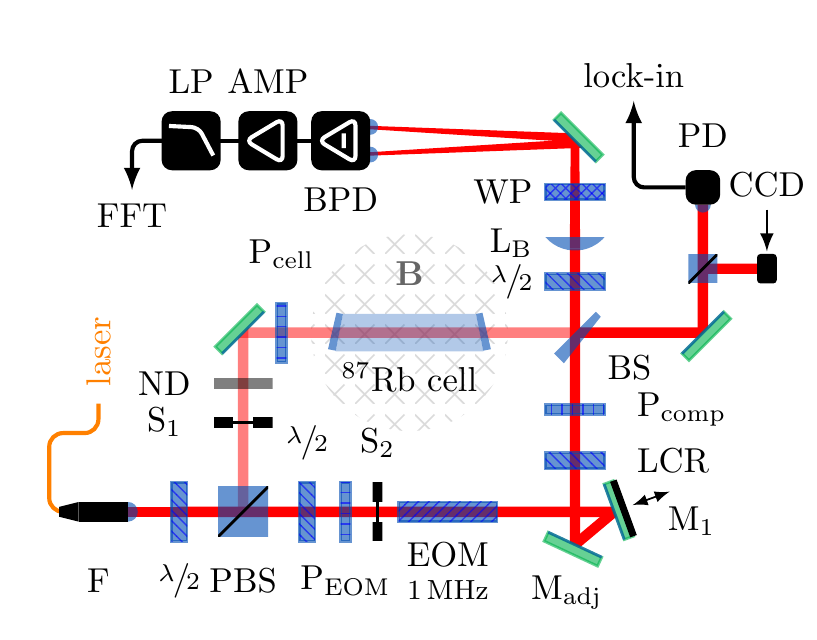}
   \caption{
   Schematic drawing of the homodyne setup with fiber output from the laser source ($\textrm{F}$), polarizing beam splitter ($\textrm{PBS}$), high extinction linear polarizer ($\textrm{P}_i$), mechanical shutter ($\textrm{S}_i$), electro-optic phase modulator ($\textrm{EOM}$), mirror with piezoelectric actuator ($\textrm{M}_1$), liquid crystal retarder ($\textrm{LCR}$), neutral density filter ($\textrm{ND}$), low wave-front distortion non-polarizing beam-splitter ($\textrm{BS}$), Wollaston prism ($\textrm{WP}$), lens ($\textrm{L}_\textrm{B}$), high speed balanced photo-receiver for SN ($\textrm{BPD}$) whose output is amplified ($\textrm{AMP}$) and low pass filtered ($\textrm{LP}$); a photo-detector for phase stabilization ($\textrm{PD}$), and imaging CCD-array for interferometer adjustment ($\textrm{CCD}$).
A solenoid underneath the ${}^{87}\textrm{Rb}$ vapor cell yields an adjustable, transverse magnetic field $B$ in order to shift the Larmor frequency (center frequency of $S_\textrm{sn}$).
}
   \label{fig:homodyne-setup}
\end{figure}
\Fref{fig:homodyne-setup} depicts the homodyne SN spectroscopy setup which is derived from a Mach-Zehnder interferometer.
A half-wave retarder and a polarizing beam splitter ($\textrm{PBS}$) split the incoming laser beam with a power ratio between the transmitted and reflected beams of $\approx 5:1$.
In this configuration, the transmitted beam forms the local oscillator path (LO) and the reflected beam the signal path.
A neutral density filter reduces the intensity in the signal path by three orders of magnitude so that the probe power at the ${}^{87}$Rb sample cell is for all presented experiments $\approx\SI[per-mode=symbol]{250}{\micro\watt\per\cm\squared}$ ($\approx\SI{8}{\micro\watt}$ for a beam of $\SI{2}{\milli\meter}$ diameter).
A linear polarizer in front of the ${}^{87}$Rb cell ensures a well defined polarization.
The laser light in the LO path is passed through a linear polarizer and an electro-optic phase modulator in order to a) precisely redefine the linear polarization and b) to modulate the phase of the electric field with \SI{1.111}{\mega\hertz} producing the error signal for the phase stabilization feedback.
A mirror mounted on a piezoelectric actuator allows to adjust the phase by up to \SI{8}{\radian} and is used to stabilize the interferometer at the desired operating point.
A liquid crystal retarder and an additional polarizer $\textrm{P}_\textrm{comp}$ allow the adjustment of the laser power in the LO arm.
Signal and LO path overlap on a wedged non-polarizing beam splitter $\textrm{BS}$ and proceed jointly to the SN measurement port (top port in \fref{fig:homodyne-setup}) and the control port (right port in \fref{fig:homodyne-setup}).
The power in the control port is sampled by a photo-detector $\textrm{PD}$ and demodulated by a lock-in amplifier which effectively detects the phase between $\mathbf{E}_\textrm{sig}$ and $\mathbf{E}_\textrm{lo}$.
Using $\textrm{M}_1$ the phase is steered by a PID loop towards the operating point, i.e., constructive interference in the measurement port with an accuracy of $\delta\varphi\approx \SI{0.13}{\radian}$.
Power measurements in both paths are realized by blocking the other path by one of two mechanical shutters $\textrm{S}_1$ or $\textrm{S}_2$ respectively, and sampling the remaining power on the photo diode $\textrm{PD}$.
The polarization alterations in the measurement port are detected by a polarization bridge consisting of a half-wave retarder, a Wollaston prism, and a high speed balanced photo-receiver.
The output of the balanced photo-receiver \footnote{Customized version of a Femto DHPCA-S with built-in photodiodes.} is post-amplified and low-pass filtered to a bandwidth of $\SI{5}{\mega\hertz}$ ($\SI{3}{\decibel}$).
The resulting voltage signal is sent to an analyzing computer that samples the voltage with a fast 12bit digital-to-analog converter operating at \SI{20}{\mega\hertz} and performs a discrete fast Fourier transform.
Please note that only electronic noise from the transimpedance amplifier in the first stage of the balanced photo-receiver has a significant influence on the experiment, whereas the electronic noise contribution of the intermediate voltage amplifier  $\SI{2.5}{\nano\volt\per\sqrt{\hertz}}$ ('post-amplification') and the digitizer  $\SI{80}{\nano\volt\per\sqrt{\hertz}}$ can be neglected.
We selected at the balanced photo-receiver two different transimpedance gains of $\SI{1e4}{\volt\per\ampere}$ and $\SI{1e5}{\volt\per\ampere}$ corresponding to noise-equivalent-power densities (NEP) of $\SI{10}{\pico\watt\per\sqrt{Hz}}$ and $\SI{2.5}{\pico\watt\per\sqrt{Hz}}$, respectively, in order to change the ratio of electronic versus optical shot noise present in the detection. Both amplifications have a bandwidth larger than the auxiliary low-pass filter of $\SI{5}{\mega\hertz}$. Thus we alter via a diametrically set amplification by the subsequent voltage amplifier the relative contribution of electronic noise with respect to spin and shot noise.

\subsection{Sample and Experimental Details}
The sample is a ${}^{87}\textrm{Rb}$ vapor cell 
with helium buffer gas and a purity of the ${}^{87}\textrm{Rb}$ isotope of $>\SI{98}{\percent}$.
The goal of the presented setup is the optical amplification of extremely small SN signals.
Therefore, the weak SN signature of the remaining ${}^{85}\textrm{Rb}$ isotope is analyzed and the measurement is carried out at room temperature where the Rb vapor partial pressure is very low.
The vapor density and optical probe power used in the presented measurements are significantly lower than the optimal values reported by \cite{lucivero_sensitivity_2017}, i.e., $n\approx\SI{1.2e10}{\per\centi\meter\squared}, P_\textrm{sig}\approx\SI{8}{\micro\watt}$. These parameters have been chosen of course on purpose in order to simulate the unfavorable conditions present in single quantum dot setups that are the main scope of our technique.

The laser source used in this setup is an external cavity diode laser stabilized by a Fizeau interferometer and the laser photon energy is blue-shifted from the ${}^{85}\textrm{Rb}$~$D_2$ resonance by $\approx{\SI{0.3}{\giga\hertz}}$.
A solenoid below the vapor cell produces a transverse magnetic field which modulates the stochastic spin orientation by the Larmor frequency $f_\textrm{L}$.
In order to separate the weak SN signature from other noise contributions, a foreground and a background SN spectrum are measured at a magnetic field of $B_\textrm{fg}=\SI{120}{\micro\tesla}$ and $B_\textrm{bg}=\SI{800}{\micro\tesla}$, respectively, and the two spectra are subtracted from each other.
All other noise contributions are independent of magnetic field and average out in the difference of the foreground and background spectra whereat only the SN contribution remains.

\section{Results}
   \begin{figure}[t]
        \centering
   \includegraphics[]{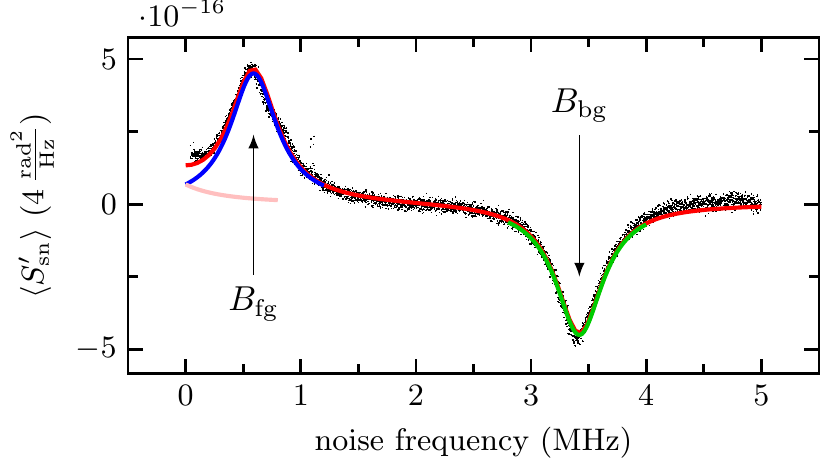}
   \caption{
   Composite average of measured SN difference spectrum of $^{85}$Rb (dots) (cf. Eq.~\ref{eq:sn-prime}) in units of Faraday rotation power spectral density.
   The solid red line shows a fit with two Lorentzians which are shifted by the Larmor frequency corresponding to $B_\textrm{fg}$ and $B_\textrm{bg}$. The field values of $B_\textrm{fg}=\SI{120}{\micro\tesla}$ and $B_\textrm{bg}=\SI{800}{\micro\tesla}$
   were chosen so both peaks fit well inside the available bandwidth. The folding of the foreground peak bears no drawback for the evaluation and nicely demonstrates the conservation of spin noise power.
    }
    \label{fig:sn-prime}
\end{figure}

\Fref{fig:sn-prime} depicts a high quality SN spectrum which is obtained by avarging over SN spectra~\footnote{The SN difference spectra exhibit often a small drift term $S_\textrm{drift}$ which arises from imperfect cancelation of the different noise sources. This drift term can be easily subtracted from the individual difference spectra. In order to guarantee the most consistent results all presented data have been corrected accordingly in the same manner.} measured at different amplifications factors $\eta_p$ and normalized by ${P^2_\textrm{sig}}$:
\begin{equation}\label{eq:sn-prime}
	S^\prime_\textrm{sn}
		= \frac{S(B=B_\textrm{fg}) - S(B=B_\textrm{bg})}{P^2_\textrm{sig}\eta_p}.
\end{equation}
The solid (red) line is a fit to the spectrum consisting of a positive (foreground, blue) and a negative (background, green) Lorentzian whereat part of the foreground Lorentzian is folded back at zero frequency (pink line).
The extremely accurate fit allows the extraction of the noise powers, the center frequencies $f_\textrm{L}$, and the full widths at half maximum $\gamma_\textrm{s}$ with high precision. Both parameters, $f_\textrm{L}$ and $\gamma_\textrm{s}$, are constant in the experiments such that the spin noise amplitude is a direct measure for optical amplification in the following evaluation of the measurements.

\begin{figure}[t]
   \centering
   \includegraphics[]{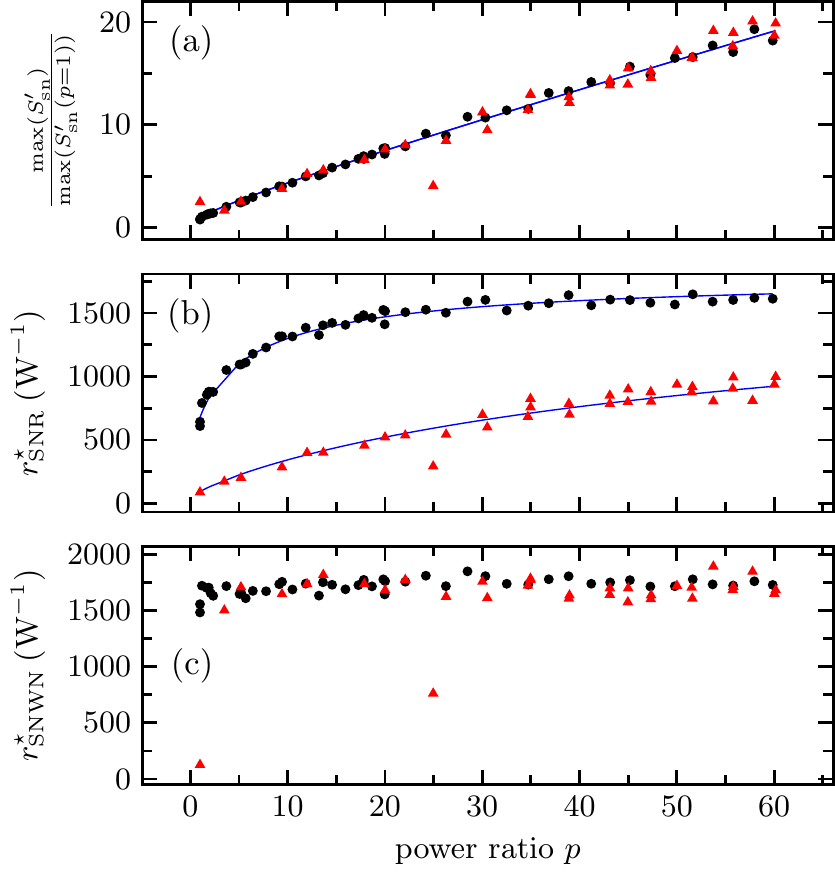}
   \caption{
   (a) Measured SN amplitude in dependence on $p$ normalized to the SN amplitude at ${p=1}$.
   The solid line is a plot of $\eta_p / \eta_1$.
   (b) Signal-to-noise ratio for weak (black circles) and large electronic noise (red triangles).
   The solid line is a fit according to \fref{eq:snr-star}.
   (c) Measured ratio of SN and photon shot noise.
    }
   \label{fig:homodyne-results}
\end{figure}

First, we study the optical amplification of the SN in dependence on the light intensity of the local oscillator. Here, we only change the power ratio $p$ but not $P_\textrm{sig}$.
\Fref{fig:homodyne-results}(a) shows the amplitude of the Lorentzian SN peak normalized to the amplitude at $p=1$ as a function of $p$.
As predicted by \fref{eq:noise-all},  the amplitude scales with ${\eta_p=(1+\sqrt{p})^2}$ (solid line).
This is a direct experimental proof of optical amplification but does not prove an improvement of the signal-to-noise ratio, yet.

All experiments shown in \fref{fig:homodyne-results} are carried out for two different scenarios, (i) high electrical amplification of the balanced receiver of $\SI{6e4}{\volt\per\watt}$ and low post-amplification of $\SI{30}{\decibel}$ (black circles) and (ii) lower electrical amplification of the balanced receiver of $\SI{6e3}{\volt\per\watt}$ and higher post-amplification of $\SI{45}{\decibel}$ (red triangles).
The typical amplifier characteristic of the balanced receiver yields for the case (i) lower electronic noise than for the case (ii), i.e., the black circles and the red triangles show measurements with different effective electronic noise contributions.
The data for the optical amplification of SN is depicted for both cases in \fref{fig:homodyne-results}(a) and shows that the optical amplification of SN is independent of the background electronic noise.

Next, we demonstrate that homodyne amplification also increases the signal-to-noise ratio $r^\star_\textrm{SNR}$ if electronic noise is the dominating noise source.
\Fref{fig:homodyne-results}(b) clearly shows the increase of $r^\star_\textrm{SNR}$ with increasing power ratio $p$.
In the case of moderate electronic noise (black circles), $r^\star_\textrm{SNR}$ increases with increasing $p$ and approaches saturation for high $p$ since white photon shot noise dominates over electronic noise, i.e. homodyne detection with non-squeezed light cannot beat the photon shot noise limit here~\cite{lucivero_squeezed-light_2016}.
In the case of higher electronic noise (red triangles), $r^\star_\textrm{SNR}$ also increases with increasing $p$ but is significantly smaller at low $p$ and does not reach saturation for the maximal $p$ in this experiment.

Finally, \fref{fig:homodyne-results}(c) shows the ratio $r^\star_\textrm{SNWN}$
which is obtained by scaling the measured SN power normalized by $P_\textrm{sig}$ with the measured optical shot noise power.
This ratio is consistently constant since the optical shot noise and the SN power increase for homodyne detection by the same proportionality factor $\eta_p$ with $P_\textrm{sig}$ staying constant. Here $r^\star_\textrm{SNWN}$ being independent of $p$ affirms the assumptions and accuracy of our model.

\section{Conclusions}
We have shown in a  proof-of-principle experiment that the resulting constraints by electronic noise can be successfully circumvented by optical homodyne amplification of spin noise in a standard spin noise spectroscopy setup and even at low frequencies.
This low frequency limit is especially important for semiconductor systems with long spin coherence times which are relevant in the framework of future semiconductor spin information processing.
Here, spin noise spectroscopy on fragile spin systems and in particular on single InGaAs quantum dots, which is usually limited by the finite electronic noise of the available balanced receivers, can be made feasible.
The experiments have been successfully carried out with a non-commercial diode laser down to SN frequencies of tenths of kHz which is sufficient for most spin noise experiments.
Nevertheless, even lower spin noise frequencies should be easily accessible utilizing stabilized ultra-narrow linewidth lasers.

\appendix

\section{An Alternative to Phase Stabilization}
An active phase stabilization loop significantly increases the setup complexity. Here, we propose an alternative scheme that explicitly averages over a changing phase $\varphi_\textrm{sig,lo}$.

The homodyne amplification has a cosine-type phase dependence which can be averaged out if the phase modulation a) acts on a different time scale than the spin dynamics and b) is faster than the characteristic drift of the setup. For a modulation much slower than the rate $f_\textrm{rec}$ at which spectra are recorded this means that the phase has to be effectively constant. For a modulation much faster than $f_\textrm{rec}$ the phase dependence must average out for each individual spectrum.
For the measured noise power density (roughly the square of \fref{eq:amp-approx}) it is easy to show that
\begin{equation}
    \label{eq:dither-approx}
    \overline{\Delta I^2} = \int_{0}^{n 2 \pi} {\Delta I}^2
        \frac{\mathrm{d}\varphi}{n 2 \pi}
        \propto I_\textrm{sig}^2 \theta_\textrm{F}^2 \left(\frac{2 + p}{2}\right) + O\left(\theta_\textrm{F}^3\right).
\end{equation}
This kind of \textit{phase dithering} reduces the effective amplification factor of the spin noise signal to $\eta_{\textrm{dith},p} = \frac{2 + p}{2}$, while keeping all other properties of the homodyne detection.
For the technical realization of this scheme the piezoelectrically actuated mirror $\textrm{M}_1$ has to be replaced ideally by a phase modulator with a very high modulation depth to ensure optimal phase averaging.

    \begin{acknowledgments}
We gratefully acknowledge financial support by the NTH school for contacts in nanosystems, the BMBF joint research project Q.com-H (16KIS0109 and 16KIS00107), and the DFG (GRK~1991, OE~177/10-1). We thank K.~Pierz from the PTB for providing the QD sample.
    \end{acknowledgments}


\begin{thebibliography}{29}%
\makeatletter
\providecommand \@ifxundefined [1]{%
 \@ifx{#1\undefined}
}%
\providecommand \@ifnum [1]{%
 \ifnum #1\expandafter \@firstoftwo
 \else \expandafter \@secondoftwo
 \fi
}%
\providecommand \@ifx [1]{%
 \ifx #1\expandafter \@firstoftwo
 \else \expandafter \@secondoftwo
 \fi
}%
\providecommand \natexlab [1]{#1}%
\providecommand \enquote  [1]{``#1''}%
\providecommand \bibnamefont  [1]{#1}%
\providecommand \bibfnamefont [1]{#1}%
\providecommand \citenamefont [1]{#1}%
\providecommand \href@noop [0]{\@secondoftwo}%
\providecommand \href [0]{\begingroup \@sanitize@url \@href}%
\providecommand \@href[1]{\@@startlink{#1}\@@href}%
\providecommand \@@href[1]{\endgroup#1\@@endlink}%
\providecommand \@sanitize@url [0]{\catcode `\\12\catcode `\$12\catcode
  `\&12\catcode `\#12\catcode `\^12\catcode `\_12\catcode `\%12\relax}%
\providecommand \@@startlink[1]{}%
\providecommand \@@endlink[0]{}%
\providecommand \url  [0]{\begingroup\@sanitize@url \@url }%
\providecommand \@url [1]{\endgroup\@href {#1}{\urlprefix }}%
\providecommand \urlprefix  [0]{URL }%
\providecommand \Eprint [0]{\href }%
\providecommand \doibase [0]{http://dx.doi.org/}%
\providecommand \selectlanguage [0]{\@gobble}%
\providecommand \bibinfo  [0]{\@secondoftwo}%
\providecommand \bibfield  [0]{\@secondoftwo}%
\providecommand \translation [1]{[#1]}%
\providecommand \BibitemOpen [0]{}%
\providecommand \bibitemStop [0]{}%
\providecommand \bibitemNoStop [0]{.\EOS\space}%
\providecommand \EOS [0]{\spacefactor3000\relax}%
\providecommand \BibitemShut  [1]{\csname bibitem#1\endcsname}%
\let\auto@bib@innerbib\@empty
\bibitem [{\citenamefont {Meier}\ and\ \citenamefont
  {Zakharchenya}(1984)}]{meier_optical_1984}%
  \BibitemOpen
  \bibinfo {editor} {\bibfnamefont {F.}~\bibnamefont {Meier}}\ and\ \bibinfo
  {editor} {\bibfnamefont {B.~P.}\ \bibnamefont {Zakharchenya}},\ eds.,\
  \href@noop {} {{\selectlanguage {english}\emph {\bibinfo {title} {Optical
  orientation}}}}\ (\bibinfo  {publisher} {North-Holland},\ \bibinfo {year}
  {1984})\BibitemShut {NoStop}%
\bibitem [{\citenamefont {Kikkawa}\ and\ \citenamefont
  {Awschalom}(1998)}]{kikkawa_resonant_1998}%
  \BibitemOpen
  \bibfield  {author} {\bibinfo {author} {\bibfnamefont {J.~M.}\ \bibnamefont
  {Kikkawa}}\ and\ \bibinfo {author} {\bibfnamefont {D.~D.}\ \bibnamefont
  {Awschalom}},\ }\href {\doibase 10.1103/PhysRevLett.80.4313} {\bibfield
  {journal} {\bibinfo  {journal} {Phys. Rev. Lett.}\ }\textbf {\bibinfo
  {volume} {80}},\ \bibinfo {pages} {4313} (\bibinfo {year}
  {1998})}\BibitemShut {NoStop}%
\bibitem [{\citenamefont {Atat\"{u}re}\ \emph {et~al.}(2007)\citenamefont
  {Atat\"{u}re}, \citenamefont {Dreiser}, \citenamefont {Badolato},\ and\
  \citenamefont {Imamoglu}}]{Atature.NatPhys.2007}%
  \BibitemOpen
  \bibfield  {author} {\bibinfo {author} {\bibfnamefont {M.}~\bibnamefont
  {Atat\"{u}re}}, \bibinfo {author} {\bibfnamefont {J.}~\bibnamefont {Dreiser}},
  \bibinfo {author} {\bibfnamefont {A.}~\bibnamefont {Badolato}}, \ and\
  \bibinfo {author} {\bibfnamefont {A.}~\bibnamefont {Imamoglu}},\ }\href
  {http://dx.doi.org/10.1038/nphys521} {\bibfield  {journal} {\bibinfo
  {journal} {Nature Phys.}\ }\textbf {\bibinfo {volume} {3}},\ \bibinfo {pages}
  {101} (\bibinfo {year} {2007})}\BibitemShut {NoStop}%
\bibitem [{\citenamefont {Dahbashi}\ \emph {et~al.}(2012)\citenamefont
  {Dahbashi}, \citenamefont {H\"{u}bner}, \citenamefont {Berski}, \citenamefont
  {Wiegand}, \citenamefont {Marie}, \citenamefont {Pierz}, \citenamefont
  {Schumacher},\ and\ \citenamefont {Oestreich}}]{Dahbashi.APL.2012}%
  \BibitemOpen
  \bibfield  {author} {\bibinfo {author} {\bibfnamefont {R.}~\bibnamefont
  {Dahbashi}}, \bibinfo {author} {\bibfnamefont {J.}~\bibnamefont {H\"{u}bner}},
  \bibinfo {author} {\bibfnamefont {F.}~\bibnamefont {Berski}}, \bibinfo
  {author} {\bibfnamefont {J.}~\bibnamefont {Wiegand}}, \bibinfo {author}
  {\bibfnamefont {X.}~\bibnamefont {Marie}}, \bibinfo {author} {\bibfnamefont
  {K.}~\bibnamefont {Pierz}}, \bibinfo {author} {\bibfnamefont {H.~W.}\
  \bibnamefont {Schumacher}}, \ and\ \bibinfo {author} {\bibfnamefont
  {M.}~\bibnamefont {Oestreich}},\ }\href {\doibase 10.1063/1.3678182}
  {\bibfield  {journal} {\bibinfo  {journal} {Appl. Phys. Lett.}\ }\textbf
  {\bibinfo {volume} {100}},\ \bibinfo {pages} {31906} (\bibinfo {year}
  {2012})}\BibitemShut {NoStop}%
\bibitem [{\citenamefont {Dahbashi}\ \emph
  {et~al.}(2014{\natexlab{a}})\citenamefont {Dahbashi}, \citenamefont
  {H\"{u}bner}, \citenamefont {Berski}, \citenamefont {Pierz},\ and\ \citenamefont
  {Oestreich}}]{Dahbashi.PRL.2014}%
  \BibitemOpen
  \bibfield  {author} {\bibinfo {author} {\bibfnamefont {R.}~\bibnamefont
  {Dahbashi}}, \bibinfo {author} {\bibfnamefont {J.}~\bibnamefont {H\"{u}bner}},
  \bibinfo {author} {\bibfnamefont {F.}~\bibnamefont {Berski}}, \bibinfo
  {author} {\bibfnamefont {K.}~\bibnamefont {Pierz}}, \ and\ \bibinfo {author}
  {\bibfnamefont {M.}~\bibnamefont {Oestreich}},\ }\href {\doibase
  10.1103/PhysRevLett.112.156601} {\bibfield  {journal} {\bibinfo  {journal}
  {Phys. Rev. Lett.}\ }\textbf {\bibinfo {volume} {112}},\ \bibinfo {pages}
  {156601} (\bibinfo {year} {2014}{\natexlab{a}})}\BibitemShut {NoStop}%
\bibitem [{\citenamefont {R\"{o}mer}\ \emph {et~al.}(2007)\citenamefont {R\"{o}mer},
  \citenamefont {H\"{u}bner},\ and\ \citenamefont {Oestreich}}]{Romer.RSI.2007}%
  \BibitemOpen
  \bibfield  {author} {\bibinfo {author} {\bibfnamefont {M.}~\bibnamefont
  {R\"{o}mer}}, \bibinfo {author} {\bibfnamefont {J.}~\bibnamefont {H\"{u}bner}}, \
  and\ \bibinfo {author} {\bibfnamefont {M.}~\bibnamefont {Oestreich}},\ }\href
  {\doibase 10.1063/1.2794059} {\bibfield  {journal} {\bibinfo  {journal} {Rev.
  Sci. Instrum.}\ }\textbf {\bibinfo {volume} {78}},\ \bibinfo {pages} {103903}
  (\bibinfo {year} {2007})}\BibitemShut {NoStop}%
\bibitem [{\citenamefont {Henn}\ \emph {et~al.}(2013)\citenamefont {Henn},
  \citenamefont {Heckel}, \citenamefont {Beck}, \citenamefont {Kiessling},
  \citenamefont {Ossau}, \citenamefont {Molenkamp}, \citenamefont {Reuter},\
  and\ \citenamefont {Wieck}}]{henn_hot_2013}%
  \BibitemOpen
  \bibfield  {author} {\bibinfo {author} {\bibfnamefont {T.}~\bibnamefont
  {Henn}}, \bibinfo {author} {\bibfnamefont {A.}~\bibnamefont {Heckel}},
  \bibinfo {author} {\bibfnamefont {M.}~\bibnamefont {Beck}}, \bibinfo {author}
  {\bibfnamefont {T.}~\bibnamefont {Kiessling}}, \bibinfo {author}
  {\bibfnamefont {W.}~\bibnamefont {Ossau}}, \bibinfo {author} {\bibfnamefont
  {L.~W.}\ \bibnamefont {Molenkamp}}, \bibinfo {author} {\bibfnamefont
  {D.}~\bibnamefont {Reuter}}, \ and\ \bibinfo {author} {\bibfnamefont {A.~D.}\
  \bibnamefont {Wieck}},\ }\href {\doibase 10.1103/PhysRevB.88.085303}
  {\bibfield  {journal} {\bibinfo  {journal} {Phys. Rev. B}\ }\textbf {\bibinfo
  {volume} {88}},\ \bibinfo {pages} {085303} (\bibinfo {year}
  {2013})}\BibitemShut {NoStop}%
\bibitem [{\citenamefont {H\"{u}bner}\ \emph {et~al.}(2014)\citenamefont
  {H\"{u}bner}, \citenamefont {Berski}, \citenamefont {Dahbashi},\ and\
  \citenamefont {Oestreich}}]{hubner_rise_2014}%
  \BibitemOpen
  \bibfield  {author} {\bibinfo {author} {\bibfnamefont {J.}~\bibnamefont
  {H\"{u}bner}}, \bibinfo {author} {\bibfnamefont {F.}~\bibnamefont {Berski}},
  \bibinfo {author} {\bibfnamefont {R.}~\bibnamefont {Dahbashi}}, \ and\
  \bibinfo {author} {\bibfnamefont {M.}~\bibnamefont {Oestreich}},\ }\href
  {\doibase 10.1002/pssb.201350291} {\bibfield  {journal} {\bibinfo  {journal}
  {physica status solidi (b)}\ }\textbf {\bibinfo {volume} {251}},\ \bibinfo
  {pages} {1824} (\bibinfo {year} {2014})}\BibitemShut {NoStop}%
\bibitem [{\citenamefont {Oestreich}\ \emph {et~al.}(2005)\citenamefont
  {Oestreich}, \citenamefont {R\"{o}mer}, \citenamefont {Haug},\ and\
  \citenamefont {H\"{a}gele}}]{oestreich_spin_2005}%
  \BibitemOpen
  \bibfield  {author} {\bibinfo {author} {\bibfnamefont {M.}~\bibnamefont
  {Oestreich}}, \bibinfo {author} {\bibfnamefont {M.}~\bibnamefont {R\"{o}mer}},
  \bibinfo {author} {\bibfnamefont {R.~J.}\ \bibnamefont {Haug}}, \ and\
  \bibinfo {author} {\bibfnamefont {D.}~\bibnamefont {H\"{a}gele}},\ }\href
  {\doibase 10.1103/PhysRevLett.95.216603} {\bibfield  {journal} {\bibinfo
  {journal} {Phys. Rev. Lett.}\ }\textbf {\bibinfo {volume} {95}},\ \bibinfo
  {pages} {216603} (\bibinfo {year} {2005})}\BibitemShut {NoStop}%
\bibitem [{\citenamefont {Dahbashi}\ \emph
  {et~al.}(2014{\natexlab{b}})\citenamefont {Dahbashi}, \citenamefont
  {H\"{u}bner}, \citenamefont {Berski}, \citenamefont {Pierz},\ and\ \citenamefont
  {Oestreich}}]{dahbashi_optical_2014}%
  \BibitemOpen
  \bibfield  {author} {\bibinfo {author} {\bibfnamefont {R.}~\bibnamefont
  {Dahbashi}}, \bibinfo {author} {\bibfnamefont {J.}~\bibnamefont {H\"{u}bner}},
  \bibinfo {author} {\bibfnamefont {F.}~\bibnamefont {Berski}}, \bibinfo
  {author} {\bibfnamefont {K.}~\bibnamefont {Pierz}}, \ and\ \bibinfo {author}
  {\bibfnamefont {M.}~\bibnamefont {Oestreich}},\ }\href {\doibase
  10.1103/PhysRevLett.112.156601} {\bibfield  {journal} {\bibinfo  {journal}
  {Phys. Rev. Lett.}\ }\textbf {\bibinfo {volume} {112}},\ \bibinfo {pages}
  {156601} (\bibinfo {year} {2014}{\natexlab{b}})}\BibitemShut {NoStop}%
\bibitem [{\citenamefont {Wiegand}\ \emph {et~al.}(2017)\citenamefont
  {Wiegand}, \citenamefont {Smirnov}, \citenamefont {H\"{u}bner}, \citenamefont
  {Glazov},\ and\ \citenamefont {Oestreich}}]{wiegand_reoccupation_2017}%
  \BibitemOpen
  \bibfield  {author} {\bibinfo {author} {\bibfnamefont {J.}~\bibnamefont
  {Wiegand}}, \bibinfo {author} {\bibfnamefont {D.~S.}\ \bibnamefont
  {Smirnov}}, \bibinfo {author} {\bibfnamefont {J.}~\bibnamefont {H\"{u}bner}},
  \bibinfo {author} {\bibfnamefont {M.~M.}\ \bibnamefont {Glazov}}, \ and\
  \bibinfo {author} {\bibfnamefont {M.}~\bibnamefont {Oestreich}},\ }\href
  {http://arxiv.org/abs/1708.01245} {\bibfield  {journal} {\bibinfo  {journal}
  {arXiv:1708.01245 [cond-mat]}\ } (\bibinfo {year} {2017})},\ \bibinfo {note}
  {arXiv: 1708.01245}\BibitemShut {NoStop}%
\bibitem [{\citenamefont {Berski}\ \emph {et~al.}(2015)\citenamefont {Berski},
  \citenamefont {H\"{u}bner}, \citenamefont {Oestreich}, \citenamefont {Ludwig},
  \citenamefont {Wieck},\ and\ \citenamefont {Glazov}}]{berski_interplay_2015}%
  \BibitemOpen
  \bibfield  {author} {\bibinfo {author} {\bibfnamefont {F.}~\bibnamefont
  {Berski}}, \bibinfo {author} {\bibfnamefont {J.}~\bibnamefont {H\"{u}bner}},
  \bibinfo {author} {\bibfnamefont {M.}~\bibnamefont {Oestreich}}, \bibinfo
  {author} {\bibfnamefont {A.}~\bibnamefont {Ludwig}}, \bibinfo {author}
  {\bibfnamefont {A.~D.}\ \bibnamefont {Wieck}}, \ and\ \bibinfo {author}
  {\bibfnamefont {M.}~\bibnamefont {Glazov}},\ }\href {\doibase
  10.1103/PhysRevLett.115.176601} {\bibfield  {journal} {\bibinfo  {journal}
  {Phys. Rev. Lett.}\ }\textbf {\bibinfo {volume} {115}},\ \bibinfo {pages}
  {176601} (\bibinfo {year} {2015})}\BibitemShut {NoStop}%
\bibitem [{\citenamefont {M\"{u}ller}\ \emph
  {et~al.}(2010{\natexlab{a}})\citenamefont {M\"{u}ller}, \citenamefont
  {Oestreich}, \citenamefont {R\"{o}mer},\ and\ \citenamefont
  {H\"{u}bner}}]{muller_semiconductor_2010}%
  \BibitemOpen
  \bibfield  {author} {\bibinfo {author} {\bibfnamefont {G.~M.}\ \bibnamefont
  {M\"{u}ller}}, \bibinfo {author} {\bibfnamefont {M.}~\bibnamefont {Oestreich}},
  \bibinfo {author} {\bibfnamefont {M.}~\bibnamefont {R\"{o}mer}}, \ and\ \bibinfo
  {author} {\bibfnamefont {J.}~\bibnamefont {H\"{u}bner}},\ }\href {\doibase
  10.1016/j.physe.2010.08.010} {\bibfield  {journal} {\bibinfo  {journal}
  {Physica E: Low-dimensional Systems and Nanostructures}\ }\textbf {\bibinfo
  {volume} {43}},\ \bibinfo {pages} {569} (\bibinfo {year}
  {2010}{\natexlab{a}})}\BibitemShut {NoStop}%
\bibitem [{\citenamefont {Horn}\ \emph {et~al.}(2011)\citenamefont {Horn},
  \citenamefont {M\"{u}ller}, \citenamefont {Rasel}, \citenamefont {Santos},
  \citenamefont {H\"{u}bner},\ and\ \citenamefont
  {Oestreich}}]{horn_spin-noise_2011}%
  \BibitemOpen
  \bibfield  {author} {\bibinfo {author} {\bibfnamefont {H.}~\bibnamefont
  {Horn}}, \bibinfo {author} {\bibfnamefont {G.~M.}\ \bibnamefont {M\"{u}ller}},
  \bibinfo {author} {\bibfnamefont {E.~M.}\ \bibnamefont {Rasel}}, \bibinfo
  {author} {\bibfnamefont {L.}~\bibnamefont {Santos}}, \bibinfo {author}
  {\bibfnamefont {J.}~\bibnamefont {H\"{u}bner}}, \ and\ \bibinfo {author}
  {\bibfnamefont {M.}~\bibnamefont {Oestreich}},\ }\href {\doibase
  10.1103/PhysRevA.84.043851} {\bibfield  {journal} {\bibinfo  {journal} {Phys.
  Rev. A}\ }\textbf {\bibinfo {volume} {84}},\ \bibinfo {pages} {043851}
  (\bibinfo {year} {2011})}\BibitemShut {NoStop}%
\bibitem [{\citenamefont {M\"{u}ller}\ \emph
  {et~al.}(2010{\natexlab{b}})\citenamefont {M\"{u}ller}, \citenamefont {R\"{o}mer},
  \citenamefont {H\"{u}bner},\ and\ \citenamefont {Oestreich}}]{Muller.PRB.2010}%
  \BibitemOpen
  \bibfield  {author} {\bibinfo {author} {\bibfnamefont {G.~M.}\ \bibnamefont
  {M\"{u}ller}}, \bibinfo {author} {\bibfnamefont {M.}~\bibnamefont {R\"{o}mer}},
  \bibinfo {author} {\bibfnamefont {J.}~\bibnamefont {H\"{u}bner}}, \ and\
  \bibinfo {author} {\bibfnamefont {M.}~\bibnamefont {Oestreich}},\ }\href
  {\doibase 10.1103/PhysRevB.81.121202} {\bibfield  {journal} {\bibinfo
  {journal} {Phys. Rev. B}\ }\textbf {\bibinfo {volume} {81}},\ \bibinfo
  {pages} {121202(R)} (\bibinfo {year} {2010}{\natexlab{b}})}\BibitemShut
  {NoStop}%
\bibitem [{\citenamefont {Berski}\ \emph {et~al.}(2013)\citenamefont {Berski},
  \citenamefont {Kuhn}, \citenamefont {Lonnemann}, \citenamefont {H\"{u}bner},\
  and\ \citenamefont {Oestreich}}]{Berski.PRL.2013}%
  \BibitemOpen
  \bibfield  {author} {\bibinfo {author} {\bibfnamefont {F.}~\bibnamefont
  {Berski}}, \bibinfo {author} {\bibfnamefont {H.}~\bibnamefont {Kuhn}},
  \bibinfo {author} {\bibfnamefont {J.~G.}\ \bibnamefont {Lonnemann}}, \bibinfo
  {author} {\bibfnamefont {J.}~\bibnamefont {H\"{u}bner}}, \ and\ \bibinfo
  {author} {\bibfnamefont {M.}~\bibnamefont {Oestreich}},\ }\href {\doibase
  10.1103/PhysRevLett.111.186602} {\bibfield  {journal} {\bibinfo  {journal}
  {Phys. Rev. Lett.}\ }\textbf {\bibinfo {volume} {111}},\ \bibinfo {pages}
  {186602} (\bibinfo {year} {2013})}\BibitemShut {NoStop}%
\bibitem [{\citenamefont {LaForge}\ and\ \citenamefont
  {Steeves}(2007)}]{laforge_noninvasive_2007}%
  \BibitemOpen
  \bibfield  {author} {\bibinfo {author} {\bibfnamefont {J.~M.}\ \bibnamefont
  {LaForge}}\ and\ \bibinfo {author} {\bibfnamefont {G.~M.}\ \bibnamefont
  {Steeves}},\ }\href {\doibase 10.1063/1.2785111} {\bibfield  {journal}
  {\bibinfo  {journal} {Applied Physics Letters}\ }\textbf {\bibinfo {volume}
  {91}},\ \bibinfo {pages} {121115} (\bibinfo {year} {2007})}\BibitemShut
  {NoStop}%
\bibitem [{\citenamefont {LaForge}\ and\ \citenamefont
  {Steeves}(2008)}]{laforge_machzehnder_2008}%
  \BibitemOpen
  \bibfield  {author} {\bibinfo {author} {\bibfnamefont {J.~M.}\ \bibnamefont
  {LaForge}}\ and\ \bibinfo {author} {\bibfnamefont {G.~M.}\ \bibnamefont
  {Steeves}},\ }\href {\doibase 10.1063/1.2948309} {\bibfield  {journal}
  {\bibinfo  {journal} {Review of Scientific Instruments}\ }\textbf {\bibinfo
  {volume} {79}},\ \bibinfo {pages} {063106} (\bibinfo {year}
  {2008})}\BibitemShut {NoStop}%
\bibitem [{\citenamefont {Cronenberger}\ and\ \citenamefont
  {Scalbert}(2016)}]{cronenberger_quantum_2016}%
  \BibitemOpen
  \bibfield  {author} {\bibinfo {author} {\bibfnamefont {S.}~\bibnamefont
  {Cronenberger}}\ and\ \bibinfo {author} {\bibfnamefont {D.}~\bibnamefont
  {Scalbert}},\ }\href {\doibase 10.1063/1.4962863} {\bibfield  {journal}
  {\bibinfo  {journal} {Review of Scientific Instruments}\ }\textbf {\bibinfo
  {volume} {87}},\ \bibinfo {pages} {093111} (\bibinfo {year}
  {2016})}\BibitemShut {NoStop}%
\bibitem [{\citenamefont {Crooker}\ \emph {et~al.}(2004)\citenamefont
  {Crooker}, \citenamefont {Rickel}, \citenamefont {Balatsky},\ and\
  \citenamefont {Smith}}]{crooker_spectroscopy_2004}%
  \BibitemOpen
  \bibfield  {author} {\bibinfo {author} {\bibfnamefont {S.~A.}\ \bibnamefont
  {Crooker}}, \bibinfo {author} {\bibfnamefont {D.~G.}\ \bibnamefont {Rickel}},
  \bibinfo {author} {\bibfnamefont {A.~V.}\ \bibnamefont {Balatsky}}, \ and\
  \bibinfo {author} {\bibfnamefont {D.~L.}\ \bibnamefont {Smith}},\ }\href
  {\doibase 10.1038/nature02804} {\bibfield  {journal} {\bibinfo  {journal}
  {Nature}\ }\textbf {\bibinfo {volume} {431}},\ \bibinfo {pages} {49}
  (\bibinfo {year} {2004})}\BibitemShut {NoStop}%
\bibitem [{\citenamefont {Ma}\ \emph {et~al.}(2016)\citenamefont {Ma},
  \citenamefont {Shi}, \citenamefont {Qian}, \citenamefont {Li},\ and\
  \citenamefont {Ji}}]{ma_spin_2016}%
  \BibitemOpen
  \bibfield  {author} {\bibinfo {author} {\bibfnamefont {J.}~\bibnamefont
  {Ma}}, \bibinfo {author} {\bibfnamefont {P.}~\bibnamefont {Shi}}, \bibinfo
  {author} {\bibfnamefont {X.}~\bibnamefont {Qian}}, \bibinfo {author}
  {\bibfnamefont {W.}~\bibnamefont {Li}}, \ and\ \bibinfo {author}
  {\bibfnamefont {Y.}~\bibnamefont {Ji}},\ }\href {\doibase
  10.1088/1674-1056/25/11/117203} {\bibfield  {journal} {\bibinfo  {journal}
  {Chinese Physics B}\ }\textbf {\bibinfo {volume} {25}},\ \bibinfo {pages}
  {117203} (\bibinfo {year} {2016})}\BibitemShut {NoStop}%
\bibitem [{\citenamefont {Loss}\ and\ \citenamefont
  {DiVincenzo}(1998)}]{loss_quantum_1998}%
  \BibitemOpen
  \bibfield  {author} {\bibinfo {author} {\bibfnamefont {D.}~\bibnamefont
  {Loss}}\ and\ \bibinfo {author} {\bibfnamefont {D.~P.}\ \bibnamefont
  {DiVincenzo}},\ }\href {\doibase 10.1103/PhysRevA.57.120} {\bibfield
  {journal} {\bibinfo  {journal} {Phys. Rev. A}\ }\textbf {\bibinfo {volume}
  {57}},\ \bibinfo {pages} {120} (\bibinfo {year} {1998})}\BibitemShut
  {NoStop}%
\bibitem [{\citenamefont {Imamoglu}\ \emph {et~al.}(1999)\citenamefont
  {Imamoglu}, \citenamefont {Awschalom}, \citenamefont {Burkard},
  \citenamefont {DiVincenzo}, \citenamefont {Loss}, \citenamefont {Sherwin},\
  and\ \citenamefont {Small}}]{imamoglu_quantum_1999}%
  \BibitemOpen
  \bibfield  {author} {\bibinfo {author} {\bibfnamefont {A.}~\bibnamefont
  {Imamoglu}}, \bibinfo {author} {\bibfnamefont {D.~D.}\ \bibnamefont
  {Awschalom}}, \bibinfo {author} {\bibfnamefont {G.}~\bibnamefont {Burkard}},
  \bibinfo {author} {\bibfnamefont {D.~P.}\ \bibnamefont {DiVincenzo}},
  \bibinfo {author} {\bibfnamefont {D.}~\bibnamefont {Loss}}, \bibinfo {author}
  {\bibfnamefont {M.}~\bibnamefont {Sherwin}}, \ and\ \bibinfo {author}
  {\bibfnamefont {A.}~\bibnamefont {Small}},\ }\href {\doibase
  10.1103/PhysRevLett.83.4204} {\bibfield  {journal} {\bibinfo  {journal}
  {Phys. Rev. Lett.}\ }\textbf {\bibinfo {volume} {83}},\ \bibinfo {pages}
  {4204} (\bibinfo {year} {1999})}\BibitemShut {NoStop}%
\bibitem [{\citenamefont {Delteil}\ \emph {et~al.}(2017)\citenamefont
  {Delteil}, \citenamefont {Sun}, \citenamefont {F\"{a}lt},\ and\ \citenamefont
  {Imamoglu}}]{delteil_realization_2017}%
  \BibitemOpen
  \bibfield  {author} {\bibinfo {author} {\bibfnamefont {A.}~\bibnamefont
  {Delteil}}, \bibinfo {author} {\bibfnamefont {Z.}~\bibnamefont {Sun}},
  \bibinfo {author} {\bibfnamefont {S.}~\bibnamefont {F\"{a}lt}}, \ and\ \bibinfo
  {author} {\bibfnamefont {A.}~\bibnamefont {Imamoglu}},\ }\href {\doibase
  10.1103/PhysRevLett.118.177401} {\bibfield  {journal} {\bibinfo  {journal}
  {Phys. Rev. Lett.}\ }\textbf {\bibinfo {volume} {118}},\ \bibinfo {pages}
  {177401} (\bibinfo {year} {2017})}\BibitemShut {NoStop}%
\bibitem [{\citenamefont {Jackson}(2007)}]{jackson_electrodynamics_2007}%
  \BibitemOpen
  \bibfield  {author} {\bibinfo {author} {\bibfnamefont {J.~D.}\ \bibnamefont
  {Jackson}},\ }in\ \href
  {http://onlinelibrary.wiley.com/doi/10.1002/9783527600441.oe014/abstract}
  {{\selectlanguage {english}\emph {\bibinfo {booktitle} {The {Optics}
  {Encyclopedia}}}}}\ (\bibinfo  {publisher} {Wiley-VCH Verlag GmbH \& Co.
  KGaA},\ \bibinfo {year} {2007})\BibitemShut {NoStop}%
\bibitem [{Note1()}]{Note1}%
  \BibitemOpen
  \bibinfo {note} {Customized version of a Femto DHPCA-S with built-in
  photodiodes.}\BibitemShut {Stop}%
\bibitem [{\citenamefont {Lucivero}\ \emph {et~al.}(2017)\citenamefont
  {Lucivero}, \citenamefont {Dimic}, \citenamefont {Kong}, \citenamefont
  {Jim\'{e}nez-Mart\'{\i}nez},\ and\ \citenamefont
  {Mitchell}}]{lucivero_sensitivity_2017}%
  \BibitemOpen
  \bibfield  {author} {\bibinfo {author} {\bibfnamefont {V.~G.}\ \bibnamefont
  {Lucivero}}, \bibinfo {author} {\bibfnamefont {A.}~\bibnamefont {Dimic}},
  \bibinfo {author} {\bibfnamefont {J.}~\bibnamefont {Kong}}, \bibinfo {author}
  {\bibfnamefont {R.}~\bibnamefont {Jim\'{e}nez-Mart\'{\i}nez}}, \ and\ \bibinfo
  {author} {\bibfnamefont {M.~W.}\ \bibnamefont {Mitchell}},\ }\href {\doibase
  10.1103/PhysRevA.95.041803} {\bibfield  {journal} {\bibinfo  {journal} {Phys.
  Rev. A}\ }\textbf {\bibinfo {volume} {95}},\ \bibinfo {pages} {041803}
  (\bibinfo {year} {2017})}\BibitemShut {NoStop}%
\bibitem [{Note2()}]{Note2}%
  \BibitemOpen
  \bibinfo {note} {The SN difference spectra exhibit often a small drift term
  $S_\protect \textrm {drift}$ which arises from imperfect cancelation of the
  different noise sources. This drift term can be easily subtracted from the
  individual difference spectra. In order to guarantee the most consistent
  results all presented data have been corrected accordingly in the same
  manner.}\BibitemShut {Stop}%
\bibitem [{\citenamefont {Lucivero}\ \emph {et~al.}(2016)\citenamefont
  {Lucivero}, \citenamefont {Jim\'{e}nez-Mart\'{\i}nez}, \citenamefont {Kong},\ and\
  \citenamefont {Mitchell}}]{lucivero_squeezed-light_2016}%
  \BibitemOpen
  \bibfield  {author} {\bibinfo {author} {\bibfnamefont {V.~G.}\ \bibnamefont
  {Lucivero}}, \bibinfo {author} {\bibfnamefont {R.}~\bibnamefont
  {Jim\'{e}nez-Mart\'{\i}nez}}, \bibinfo {author} {\bibfnamefont {J.}~\bibnamefont
  {Kong}}, \ and\ \bibinfo {author} {\bibfnamefont {M.~W.}\ \bibnamefont
  {Mitchell}},\ }\href {\doibase 10.1103/PhysRevA.93.053802} {\bibfield
  {journal} {\bibinfo  {journal} {Phys. Rev. A}\ }\textbf {\bibinfo {volume}
  {93}},\ \bibinfo {pages} {053802} (\bibinfo {year} {2016})}\BibitemShut
  {NoStop}%
\end{thebibliography}
\end{document}